\documentclass[epj,nopacs]{svjour}

\usepackage{amssymb}
\usepackage{epsfig}
\usepackage{eucal}
\usepackage{amsmath}

\newcommand{\beq}{\begin{equation}}
\newcommand{\eeq}{\end{equation}}
\newcommand{\bea}{\begin{eqnarray}}
\newcommand{\eea}{\end{eqnarray}}
\newcommand{\non}{\nonumber}
\newcommand{\eq}[1]{eq.~(\ref{#1})}         

\newcommand{\e}{\varepsilon} 
\newcommand{\be}{\beta} 

\headnote{
\hfill \parbox{4cm}{\tt \normalsize CERN-TH/2002-082 \\ TTP02-03 \\}}

\title{The radiative return at small angles: virtual corrections}

\author{Johann H. K\"uhn\inst{1,2}\thanks{\email{jk@particle.uni-karlsruhe.de}}
\and Germ\'an Rodrigo\inst{1}\thanks{Supported in part by 
E.U. TMR grant HPMF-CT-2000-00989; \email{german.rodrigo@cern.ch}} 
}

\institute{Theory Division, CERN, CH-1211 Geneva 23, Switzerland. \and
Institut f\"ur Theoretische Teilchenphysik,
Universit\"at Karlsruhe, D-76128 Karlsruhe, Germany.}

\date{Received: April 20, 2002}

\abstract{Virtual corrections for electron--positron annihilation into 
one real and one off-shell photon of invariant mass $Q^2$ are evaluated.
Special attention is paid to those configurations where the real photon 
is collinear with the beam direction. This calculation is an important
ingredient of a Monte Carlo program, which simulates events with tagged 
photons from initial-state radiation, including NLO corrections.}

\begin{document}

\authorrunning{}
\titlerunning{The radiative return at small angles}
\maketitle

\section{Introduction}

The total cross section for electron--positron annihilation into
hadrons is one of the fundamental observables in particle physics. Its
high energy behaviour provides one of the first and still most
convincing arguments for the point-like nature of quarks. Its
normalization was evidence for the existence of quarks of three
different colours, and the recent, precise measurements even allow
for an excellent determination of the strong coupling at very high 
\cite{:2001xv} and intermediate energies 
(e.g. \cite{Kuhn:2001dm} and refs. therein) 
through the influence of QCD corrections.

Weighted integrals over the cross section with properly chosen kernels
are, furthermore, a decisive input for electroweak precision tests.
This applies, for example, to the electromagnetic coupling at higher
energies or to the anomalous magnetic moment of the muon.

Of particular importance for these two applications is the low energy region,
say from threshold up to centre-of-mass (cms) energies of approximately 
3~GeV and 10~GeV, respectively. Recent measurements based on energy scans 
between 2 and 5~GeV have improved the accuracy in part of this
range. However, similar, or even further improvements below 2~GeV
would be highly welcome. The region between 1.4~GeV and 2~GeV, in
particular, is poorly studied and no collider will cover this region 
in the near future. Improvements or even an independent cross-check 
of the precise measurements of the pion form factor in the low 
energy region by the CMD2 and DM2 collaborations 
would be extremely useful, since this dominates in
the analysis of the muon anomalous magnetic moment.

Experiments at present electron--positron colliders operate mostly at 
fixed energies, albeit with enormous luminosity, with BaBar
and BELLE at 10.6~GeV and KLOE at 1.02 GeV as most prominent 
examples. 

This peculiar feature allows the use of the radiative return, i.e. the reaction
\begin{equation}
e^+(p_1)+ e^- (p_2) \to \gamma(k) + \gamma^*(Q) (\to \mathrm{hadrons})~,
\label{eq:reaction}
\end{equation}
to explore a wide range of $Q^2$ in a single 
experiment~\cite{Binner:1999bt,Melnikov:2000gs,Czyz:2000wh,Spagnolo:1999mt,Khoze:2001fs,Hoefer:2001mx,Khoze:2002ix}. 

Nominally masses of the hadronic system between $2 m_\pi$ and the cms
energy of the experiment are accessible. In practice it is useful to
consider only events with a hard photon --- tagged or untagged ---
to clearly identify  the reaction, which lowers the energy 
significantly.

The study  of events with photons emitted under both large and small 
angles, and thus at a significantly enhanced rate, is particularly 
attractive for the $\pi^+\pi^-$ final state with its clear signature, 
an investigation performed at present at 
DA$\Phi$NE~\cite{Aloisio:2001xq,Denig:2001ra,Adinolfi:2000fv}.
Events with a tagged photon, emitted under a large angle with
respect to the beam, have a clear signature and are thus particularly
suited to the analysis of hadronic final states of higher
multiplicity~\cite{babar}.

The inclusion of radiative corrections is essential for the precise
extraction of the cross section, which is necessarily based on a Monte
Carlo simulation. A first program, called EVA, was constructed
some time ago~\cite{Binner:1999bt}. 
It includes initial-state radiation (ISR), final-state
radiation (FSR), their interference, and the dominant radiative
corrections from additional collinear radiation 
through structure function techniques~\cite{Caffo:1997yy}.

The complete NLO corrections have recently been implemented in a
program called PHOKHARA~\cite{Rodrigo:2001kf}. Both programs, 
however, were designed to simulate reactions with 
tagged photons, i.e. at least one photon was required to be emitted 
under large angles. 

An important ingredient in the extension of the NLO Monte Carlo
program PHOKHARA to small photon angles is the evaluation of the virtual
corrections to the reaction (\ref{eq:reaction}) in the limit $m_e^2/s \ll 1$,
which are equally valid for large and small angles.  
Compact results for the one-loop two-, three- and four-point functions 
that enter this calculation can be found in the 
literature~\cite{'tHooft:1978xw,Beenakker:1988jr} for arbitrary
values of $m_e^2/s$. However, the combination of these analytical
expressions with the relevant coefficients is numerically unstable in
the limit of small mass and angles. A compact, numerically stable result, 
valid for an arbitrarily small photon angle, is therefore required.
As a consequence of the highly singular kinematic coefficients, terms
proportional to $m_e^2$ and even $m_e^4$  must be kept in the expansion, 
which after angular integration will contribute to the total cross 
section even in the limit $m_e^2/s\to0$.  

The present paper extends the analysis of Ref.~\cite{Rodrigo:2001jr} 
where the corrections from virtual and soft photons were presented for 
the case of large angles. In section 2 we recall the basic
definitions and describe the systematic procedure used in the
expansion of the results for small $m_e^2/s$ and small angles 
simultaneously. The analytic results for real and imaginary parts of the
leptonic tensor, expressed in an angular momentum basis,
are presented in section 3 and compared with 
results for related quantities that can be found in the literature.
After summation over the polarizations of the virtual photons, our
result agrees with the one of Berends, Burgers and van 
Neerven~\cite{Berends:1986yy,Berends:1988ab}.
The result of Kuraev, Merenkov and Fadin~\cite{Kuraev:cg1}
for the real part of the tensor, which was obtained for virtual Compton 
scattering $\gamma^*+e^-\to \gamma+e^-$ is related to our case by proper
analytic continuation; and indeed after analytic continuation we find
agreement for the real part\footnote{We disagree, however, with eq.(30) 
of the translated version~\cite{Kuraev:cg2} which contains a missprint.}.
Section 4 contains our summary and the conclusions.
The mass-dependent terms proportional to $m_e^2$ and $m_e^4$, expressed 
in the Cartesian basis, are given in Appendix~\ref{app:cartesian}.
The scalar loop integrals needed for the calculation are listed in 
Appendix~\ref{loopintegrals}.

\section{The leptonic tensor for the radiative return}

Consider the $e^+ e^-$ annihilation process
\begin{align}
e^+(p_1) + e^-(p_2) \rightarrow & \gamma^*(Q) + \gamma(k_1)~,
\end{align}
where the virtual photon decays into a hadronic final state,
$\gamma^*(Q) \to$ hadrons, and the real one is emitted from the initial 
state. The differential rate can be cast into the product of a leptonic 
and a hadronic tensor and the corresponding factorized phase space
\begin{equation}
d\sigma = \frac{1}{2s} L_{\mu \nu} H^{\mu \nu}
d \Phi_2(p_1,p_2;Q,k_1) d \Phi_n(Q;q_1,\cdot,q_n) 
\frac{dQ^2}{2\pi}~,
\end{equation}
where $d \Phi_n(Q;q_1,\cdot,q_n)$ denotes the hadronic 
$n$-body phase space, including all the statistical factors 
coming from the hadro\-nic final state. 

\begin{figure}
\begin{center}
\epsfig{file=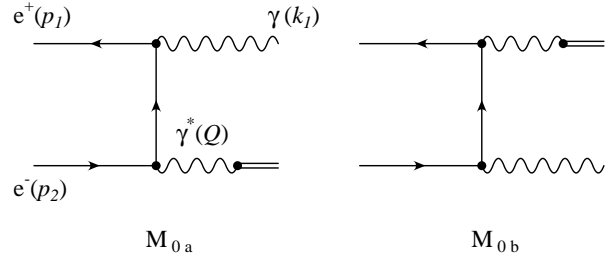,width=8cm}
\end{center}
\caption{Initial-state radiation in the annihilation 
process $e^+ e^- \rightarrow \gamma +$ hadrons at the Born level.}
\label{fig:born}
\end{figure}

\begin{figure}
\begin{center}
\epsfig{file=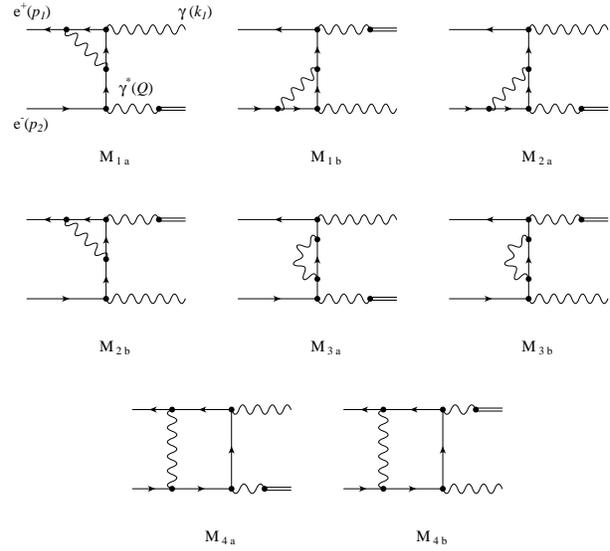,width=8cm}
\end{center}
\caption{One-loop corrections to initial-state radiation in 
$e^+ e^-$ annihilation with the emission of a virtual photon.}
\label{fig:oneloop}
\end{figure}

For an arbitrary hadronic final state, the matrix element for 
the diagrams in Fig.~\ref{fig:born} can be written 
\begin{align}
M_0 &= M_0^{\mu} J_{\mu}~, \non 
\end{align}
where $J_{\mu}$ is the hadronic current and $M_0^{\mu}$ is the 
leptonic current in lowest order.
Summing over the polarizations of the real photon,
averaging over the polarizations of the initial $e^+ e^-$ state,
and using current conservation, $Q_{\mu} J^{\mu} = 0$, 
the leptonic tensor
\begin{equation*}
L_0^{\mu \nu} = \overline{M_0^{\mu} M_0^{\nu +}}
\end{equation*}
can be written in the following form: 
\begin{align}
L_0^{\mu \nu} &= 
\frac{(4 \pi \alpha/s)^2}{q^4} \; \bigg[ \left( 
\frac{2 m^2 q^2(1-q^2)^2}{y_1^2 y_2^2}
- \frac{2 q^2+y_1^2+y_2^2}{y_1 y_2} \right) g^{\mu \nu} \non \\ & 
+ \left(\frac{8 m^2}{y_2^2} - \frac{4q^2}{y_1 y_2} \right) 
\frac{p_1^{\mu} p_1^{\nu}}{s} 
+ \left(\frac{8 m^2}{y_1^2} - \frac{4q^2}{y_1 y_2} \right) 
\frac{p_2^{\mu} p_2^{\nu}}{s} \non \\
& - \left( \frac{8 m^2}{y_1 y_2} \right) 
\frac{p_1^{\mu} p_2^{\nu} + p_1^{\nu} p_2^{\mu}}{s} \bigg]~, 
\label{Lmunu0}
\end{align}
with 
\begin{equation}
y_i = \frac{2 k_1 \cdot p_i}{s}~, 
\qquad m^2=\frac{m_e^2}{s}~, \qquad q^2=\frac{Q^2}{s}~.
\label{dimensionless}
\end{equation}
It is symmetric under the exchange of the electron and 
the positron momenta. Expressing the bilinear products $y_i$ 
by the photon emission angle in the cms frame
\begin{equation*}
y_{1,2} = \frac{1-q^2}{2}(1 \mp \be \cos \theta)~, 
\qquad \be = \sqrt{1-4m^2}~,
\end{equation*}
and rewriting the two-body phase space  
\begin{equation}
d \Phi_2(p_1,p_2;Q,k_1) = \frac{1-q^2}{32 \pi^2} d \Omega~,
\end{equation}
it is evident that expression (\ref{Lmunu0}) contains several 
singularities: soft singularities for $q^2\rightarrow 1$ and 
collinear singularities for $\cos \theta \rightarrow \pm 1$.
The former are avoided by requiring a minimal photon energy.
The latter are regulated by the electron mass.

The physics of the hadronic system, whose description is model-dependent, 
enters only through the hadronic tensor 
\begin{equation}
H^{\mu \nu} = J^{\mu} J^{\nu +}~,
\end{equation}
where the hadronic current has to be parametrized through form factors.
For two charged pions in the final state, the current 
\begin{equation}
J^{\mu}_{2\pi} = i e F_{2\pi}(Q^2) \; (q_{\pi^+}-q_{\pi^-})^{\mu}~,
\end{equation}
where $q_{\pi^+}$ and $q_{\pi^-}$ are the momenta of the $\pi^+$ and
$\pi^-$ respectively, is determined by only one function,
the pion form factor $F_{2\pi}$~\cite{Kuhn:1990ad}.
The hadronic current for four pions exhibits a more complicated structure
and has been discussed in~\cite{Czyz:2000wh,Decker:1994af,Ecker:2002cw}.

At NLO, the leptonic tensor receives contributions 
both from one-loop corrections 
arising from the insertion of virtual photon lines in the tree 
diagrams of Fig.~\ref{fig:born} and from
the emission of an extra real photon from the initial state. 
In this paper, we consider only the emission of soft photons. 
The implementation of these results in the program PHOKHARA 
and the discussion of their physical consequences 
will be considered in a separate work~\cite{inpreparation}.

At NLO, the leptonic tensor has the following general form\footnote{The 
present convention differs from the one in Ref.~\cite{Rodrigo:2001jr} 
by a factor $1/(y_1 y_2)$.}: 
\begin{align}
L^{\mu \nu} &=  \frac{(4 \pi \alpha/s)^2}{q^4} \;
\bigg[ a_{00} \; g^{\mu \nu} + a_{11} \; \frac{p_1^{\mu} p_1^{\nu}}{s}
 + a_{22} \; \frac{p_2^{\mu} p_2^{\nu}}{s} \non \\
&+ a_{12} \; \frac{p_1^{\mu} p_2^{\nu} + p_2^{\mu} p_1^{\nu}}{s}
+ i \pi \; a_{-1} \; 
\frac{p_1^{\mu} p_2^{\nu} - p_2^{\mu} p_1^{\nu}}{s} \bigg]~. 
\label{generaltensor}
\end{align}
Terms proportional to $Q^{\mu}$ are absent as a consequence of 
current conservation. The scalar coefficients $a_{ij}$ and $a_{-1}$ 
allow the following expansion 
\begin{equation}
a_{ij} = a_{ij}^{(0)} + \frac{\alpha}{\pi} \; a_{ij}^{(1)}~, \qquad
a_{-1} = \frac{\alpha}{\pi} \; a_{-1}^{(1)}~.
\end{equation}
The LO coefficients $a_{ij}^{(0)}$ can be directly read from~\eq{Lmunu0}
\begin{align}
a_{00}^{(0)} &= \frac{2 m^2 q^2(1-q^2)^2}{y_1^2 y_2^2}
- \frac{2 q^2+y_1^2+y_2^2}{y_1 y_2}~, \non \\
a_{11}^{(0)} &= \frac{8 m^2}{y_2^2} - \frac{4q^2}{y_1 y_2}~, \qquad
a_{22}^{(0)} = a_{11}^{(0)} (y_1 \leftrightarrow y_2)~, \non \\
a_{12}^{(0)} &= - \frac{8 m^2}{y_1 y_2} ~. 
\end{align}
The imaginary antisymmetric piece proportional to $a_{-1}$ appears for 
the first time at NLO. The leptonic tensor therefore remains fully 
symmetric only at LO. 

As an alternative one can replace the Cartesian basis (\eq{generaltensor})
by a basis derived from the three circular polarization vectors
of the virtual photon $\varepsilon_L$ and $\varepsilon_\pm$,
defined through
\begin{align}
\varepsilon_L^{\mu} = &
\frac{2 \left(q^{\mu} q^{\nu} - g^{\mu \nu} q^2 \right)}
{\sqrt{s \; q^2}(1-q^2)} k_{1\nu}~, \non \\
\varepsilon_1^{\mu} = & 
\frac{2 \left(q^{\mu} q^{\nu} - g^{\mu \nu} q^2 \right)}
{\sqrt{s} \; q^2(1-q^2)^2 \beta \sin \theta}~, \non \\
& \times \left( (y_1-q^2 y_2) p_{1\nu}
 - (y_2-q^2 y_1) p_{2\nu} \right)~, \non \\
\varepsilon_2^{\mu} = & 
\frac{2 \left(q^{\mu} q^{\nu} - g^{\mu \nu} q^2 \right)}
{\sqrt{s^3} \; q^2(1-q^2) \beta \sin \theta} \; 
\epsilon_{\nu \eta \rho \sigma} 
k_1^{\eta} p_1^{\rho} p_2^{\sigma}~, \non \\
\varepsilon_\pm^{\mu} = & \frac{1}{\sqrt{2}}
\left( \varepsilon_1^{\mu} \pm i \varepsilon_{2}^{\mu} \right)~,
\label{circularbasis}
\end{align}
where $\epsilon_{\nu \eta \rho \sigma}$ is the four-dimensional 
totally antisymmetric Levi-Civita tensor, $\epsilon_{0123}=-1$.
The leptonic tensor is thus given by 
\begin{align}
L^{\mu \nu} &=  \frac{(4 \pi \alpha/s)^2}{q^4} \;
\sum a_{ij} \; \varepsilon_i^{*\mu} \varepsilon_j^{\nu}~, \qquad
i,j=L,\pm~.
\end{align}
Only four of the scalar coefficients are independent 
\begin{align}
 a_{L -} = a_{L +}~, \qquad 
& a_{- L} = a_{+ L} = a_{L +}^*~, \non \\
 a_{--} = a_{+ +}~, \qquad 
& a_{- +} = a_{+ -}~. \non 
\end{align}

The trace of the leptonic tensor
\begin{align}
L^{\mu \nu} (q_{\mu} q_{\nu}-g_{\mu \nu} q^2) &= 
\frac{(4 \pi \alpha/s)^2}{q^2} ( a_{LL} + 2a_{+ +} )
\end{align}
is related to the cross section after angular averaging of the hadronic
tensor. 

The relations between the components in the Cartesian and the circular
basis read as follows:
\begin{align}
a_{LL} &= - a_{00} + \frac{1}{4q^2(1-q^2)^2} 
\big[ (y_2-q^2 y_1)^2 a_{11} \non \\ &
+ (y_1-q^2 y_2)^2 a_{22} + 2 (y_1-q^2 y_2)(y_2-q^2 y_1) a_{12}  \big]~,
\non \\ 
a_{L +} &= \frac{\beta \sin \theta}{4 \sqrt{2 q^2} (1-q^2)}
\big[ (y_2-q^2 y_1) a_{11} - (y_1-q^2 y_2) a_{22}  
\non \\ & + (1+q^2)(y_1-y_2) a_{12} - i \pi a_{-1} \big]~,
\non \\ 
a_{+ -} &= \frac{\beta^2 \sin^2 \theta}{8} 
(a_{11}+a_{22}-2a_{12})~, \non \\
a_{+ +} &= a_{+-} - a_{00}~.
\end{align}
Conversely 
\begin{align}
a_{00} &= a_{+ -} - a_{+ +}~, \non \\
a_{11} &= \frac{4}{(1-q^2)^2}  
\bigg[ q^2(a_{LL}+a_{+ -}-a_{+ +})~, \non \\ &
+ \frac{2\sqrt{2q^2}(y_1-q^2 y_2)}{(1-q^2) \beta \sin \theta} \; 
\mathrm{Re} (a_{L +}) \non \\ &
+ \frac{2(y_1- q^2 y_2)^2}{(1-q^2)^2 \beta^2 \sin^2 \theta} \; a_{+ -} 
\bigg]~, \non \\
a_{22} &= a_{11}(y_1 \leftrightarrow y_2)~, \qquad
a_{12} = \frac{a_{11}+a_{22}}{2}-\frac{4 \; a_{+ -}}{\beta^2\sin^2\theta}~,
\non \\
a_{-1} &= - \frac{4\sqrt{2q^2}}{\pi (1-q^2) \beta \sin \theta} \; 
\mathrm{Im} (a_{L +})~.
\end{align}
The scalar coefficients in the circular basis are given at LO by 
\begin{align}
a^{(0)}_{LL} &= \frac{2q^2\beta^2 \sin^2 \theta}{y_1 y_2}~, \non \\
a^{(0)}_{L+} &= \frac{\sqrt{q^2}(y_1-y_2)\beta \sin \theta}
{\sqrt{2}(1-q^2)y_1 y_2} \left(1+q^2-\frac{2m^2(1-q^2)^2}{y_1 y_2} \right)~, 
\non \\
a^{(0)}_{+-} &= \frac{\beta^2 \sin^2 \theta}{y_1 y_2}
\left( -q^2 + \frac{m^2(1-q^2)^2}{y_1 y_2} \right)~, \non \\
a^{(0)}_{++} &= a^{(0)}_{+-} + \frac{2q^2+y_1^2+y_2^2}{y_1 y_2}
- \frac{2m^2q^2(1-q^2)^2}{y_1^2 y_2^2}~.
\end{align}

The one-loop matrix elements (Fig.~\ref{fig:oneloop})
contribute to the leptonic tensor
through their interference with the lowest order diagrams
(Fig.~\ref{fig:born}).
They contain ultraviolet (UV) and infrared (IR) divergences,
which are regularized using dimensional regularization
in $D=4-2\e$ dimensions.
The UV divergences are renormalized in the on-shell scheme.
The IR divergences are cancelled by adding the contribution
of an extra soft photon emitted from the initial state
and integrated in the phase space up to an energy cutoff
$E_{\gamma}<w\sqrt{s}$ far below $\sqrt{s}$. The result, 
which is finite, depends, however, on this soft photon cutoff.
Only the contribution from hard photons with energy 
$E_{\gamma}>w\sqrt{s}$ would cancel this dependence.

The algebraic manipulations have been carried out with the 
help of the {\it FeynCalc} Mathematica package~\cite{Mertig:1991an}. 
Using standard techniques~\cite{Passarino:1979jh}, it
automatically reduces the evaluation of the one-loop contribution to 
the calculation of a few scalar one-loop integrals
and performs the Dirac algebra. 

Since we consider the small angular region, mass terms proportional 
to $y_i^{-2}$ and even $y_i^{-3}$ arise.
Terms proportional to $m^2$ and even $m^4$ must be kept, if they 
are multiplied by $y_{i}^{-2}$ and $y_i^{-3}$ respectively. 
In the expansion of the one-loop integrals, functions that 
depend on the ratio $m^2/y_i$ cannot be expanded, in contrast to 
functions of $m^2$ or $y_i$ separately. To arrive at a systematic
approach we therefore make the replacements $m^2 \rightarrow \lambda m^2$,
$y_i \rightarrow \lambda y_i$, perform the expansion for small 
$\lambda$ up to the appropriate order, and set $\lambda=1$ at the 
end.

\section{The NLO leptonic tensor}

Combining the one-loop and the soft contribution we now arrive at the 
leptonic tensor in NLO. It will be convenient to split the 
coefficients $a^{(1)}_{ij}$ into a part that contributes at large 
angles and a part proportional to $m^2$ and $m^4$,
denoted by $a^{(1,0)}_{ij}$ and $a^{(1,m)}_{ij}$ respectively:
\begin{align}
a^{(1)}_{ij} &= a_{ij}^{(0)} \bigg[ -\log(4w^2) [1+\log(m^2)] \non \\ & 
-\frac{3}{2} \log(\frac{m^2}{q^2}) 
- 2 + \frac{\pi^2}{3} \bigg] 
+ a^{(1,0)}_{ij}+a^{(1,m)}_{ij}~.
\end{align}
The factor proportional to the LO coefficients $a^{(0)}_{ij}$ 
contains the usual soft and collinear logarithms.
The expressions are particularly
compact in the circular basis. For completeness we also repeat
the results for $a^{(1,0)}_{ij}$, which can be found 
in~\cite{Rodrigo:2001jr}\footnote{The result for the imaginary part
of $a_{L+}$ differs from $a_{-1}$ in the original version 
of~\cite{Rodrigo:2001jr}.}, albeit in the Cartesian basis:          
\begin{align}
a_{LL}^{(1,0)} &= \frac{2q^2}{(1-q^2)^2} \bigg\{ -\frac{y_1}{1-y_1}
-2\log(q^2) +2L(y_1) \non \\ & + 
\bigg[1-\frac{1-q^2}{1-y_2}-\frac{q^2}{(1-y_2)^2}\bigg]
\log(\frac{y_1}{q^2})+[y_1\leftrightarrow y_2] \bigg\}~, \non \\
\end{align}
\begin{align}
a_{L+}^{(1,0)} &= \frac{\beta \sin \theta}{4 \sqrt{2 q^2} (1-q^2)} \bigg\{
\frac{y_1(1+q^2)^2}{(1-y_1)(1-y_2)} - \frac{(1-q^2)^2}{y_2} \non \\ & 
-\frac{2q^2(2+q^2)}{y_2} \log(q^2)
+\bigg[ 2(1-q^2) \bigg( \frac{1}{y_1} -\frac{q^2}{y_2} \bigg)
\non \\ & -\frac{2(1+q^2)}{1-y_2} - \frac{q^2(1+q^2)^2}{(1-y_2)^2}
\bigg] \log(\frac{y_1}{q^2}) \non \\ &
-2q^2 \bigg[ \frac{1+2q^2}{y_1} + \frac{y_1}{y_2^2} \bigg] L(y_1) 
- \frac{i \pi q^2}{y_1 y_2} \bigg[ \frac{2\log(1-y_1)}{y_1} \non \\ &
+ \frac{1-q^2}{1-y_1} + \frac{q^2}{(1-y_1)^2} \bigg] 
 - [y_1\leftrightarrow y_2] \bigg\}~,
\end{align}
\begin{align}
a_{++}^{(1,0)} &= \frac{1}{2} \bigg\{ \frac{1-y_2}{y_1} + 
\frac{1}{(1-q^2)^2} \bigg(
\frac{1+q^4}{1-y_1} - 2 + 2 \log(q^2) \bigg)
+\non \\ & + \frac{1}{1-q^2}
\bigg[ \frac{1+q^4}{1-q^2} \bigg( \frac{q^2}{(1-y_2)^2} - 1 \bigg) 
+\frac{3-q^4}{1-y_2} \bigg] \log(\frac{y_1}{q^2}) \non \\ &
+\frac{2}{1-q^2} \bigg[ \frac{(1+q^4)y_2}{(1-q^2)y_1}
+\frac{2}{y_2}\bigg] L(y_1) +[y_1\leftrightarrow y_2] \bigg\}~,
\end{align}
\begin{align}
a_{+-}^{(1,0)} &= \frac{q^2}{(1-q^2)^2} \bigg\{\frac{1+q^4}{2q^2(1-y_1)}
- \frac{1}{q^2} - \frac{y_2}{y_1} \log(q^2) \non \\ &
- \bigg[ 1 +\frac{(1-q^2)^2}{y_2} -\frac{q^2(1-q^2)}{1-y_2} 
- \frac{1+q^4}{2(1-y_2)^2}
 \bigg] \non \\ & \times \log(\frac{y_1}{q^2}) 
- \bigg[ 2 + \frac{(1-q^2)^2}{y_2^2} \bigg] L(y_1)
+[y_1\leftrightarrow y_2] \bigg\}~,
\end{align}
with
\begin{align*}
L(y_i) = Li_2(\frac{-y_i}{q^2})-Li_2(1-\frac{1}{q^2}) + 
\log(q^2+y_i) \log(\frac{y_i}{q^2})~, 
\end{align*}
where $Li_2$ is the Spence or dilogarithm function. The result in the 
Cartesian basis has been given in Ref.~\cite{Rodrigo:2001jr}. The terms 
proportional to powers of $m^2$ are given by
\begin{align}
a_{LL}^{(1,m)} = 0~,
\end{align}
\begin{align}
a_{L+}^{(1,m)} &= \frac{\sqrt{q^2} \beta \sin \theta}{4 \sqrt{2}} \bigg\{ 
 \frac{4m^2}{y_1^2} \bigg[ \log(q^2) \log(\frac{y_1^4}{m^4 q^2}) \non \\ & 
+ 4 Li_2(1-q^2) 
+ \frac{3}{2} \bigg( Li_2(1-\frac{y_1}{m^2})-\frac{\pi^2}{6} \bigg) \bigg] 
\non \\ & - n(y_1,\frac{2(1-3q^2)}{q^2}) 
+\frac{2m^2  N(y_1)}{y_1(m^2(1-q^2)-y_1)} \non \\ &
- [y_1 \leftrightarrow y_2] \bigg\}~,
\end{align}
\begin{align}
a_{++}^{(1,m)} &= - \frac{m^2 q^2}{y_1^2} \bigg[
\log(q^2) \log(\frac{y_1^4}{m^4 q^2}) + 4 Li_2(1-q^2) \non \\ &
+ Li_2(1-\frac{y_1}{m^2}) - \frac{\pi^2}{6}  \bigg] 
- \frac{m^2(1-q^2)}{y_1^2} \bigg[ 1 - \log(\frac{y_1}{m^2}) \non \\ &
+ \frac{m^2}{y_1} \bigg( Li_2(1-\frac{y_1}{m^2})-\frac{\pi^2}{6} \bigg) 
\bigg] + \frac{q^2}{2} n(y_1,\frac{1-3q^2}{q^2}) \non \\ &
+ [y_1 \leftrightarrow y_2]~,
\label{apiupiu}
\end{align}
and
\begin{align}
 a_{+-}^{(1,m)} &= \frac{\beta^2 \sin^2 \theta}{8} \bigg\{
 \frac{4m^2}{y_1^2} \bigg[ \log(q^2) \log(\frac{y_1^4}{m^4 q^2}) 
+ 4 Li_2(1-q^2) \non \\ & 
+ 2 \bigg( Li_2(1-\frac{y_1}{m^2})-\frac{\pi^2}{6} \bigg) \bigg] 
- \frac{1-q^2}{q^2} n(y_1,\frac{3-7q^2}{1-q^2}) \non \\ &
+\frac{2m^2 (1-q^2)}{y_1(m^2(1-q^2)-y_1)} \bigg[
\frac{1}{q^2} \log(\frac{y_1}{m^2}) + \frac{\log(q^2)}{1-q^2} \non \\ &
+ \bigg(\frac{3-q^2}{1-q^2}
+ \frac{m^2}{m^2(1-q^2)-y_1}\bigg) N(y_1) \bigg] \non \\ &
+ [y_1 \leftrightarrow y_2] \bigg\}~.
\end{align}
The coefficient $a_{L+}$ is antisymmetric with respect to the exchange 
$[y_1 \leftrightarrow y_2]$, while all the others are symmetric.
Only $a_{L+}^{(1,0)}$ contributes to the imaginary part.
Notice that the mass-suppressed terms are all real.
The functions $n(y_i,z)$ and $N(y_i)$ are defined through
\begin{align}
n(y_i,z) & = \frac{m^2}{y_i(m^2-y_i)} \bigg[ 1 + z \; \log(\frac{y_i}{m^2}) 
\bigg] \non \\ &
+ \frac{m^2}{(m^2-y_i)^2} \log(\frac{y_i}{m^2})~,
\end{align}
\begin{align}
N(y_i) &= \log(q^2) \log(\frac{y_i}{m^2})+Li_2(1-q^2) \non \\ & +
Li_2(1-\frac{y_i}{m^2})-\frac{\pi^2}{6}~.
\end{align}
The apparent singularity of $n$ inside the limits of phase space is 
compensated by the zero in the numerator. For the numerical 
evaluation in the region $y_i$ close to $m^2$ we use
\begin{align}
n(y_i,z) \big|_{y_i \to m^2}&=
\frac{1}{y_i} \bigg[ 1 + z \; \log(\frac{y_i}{m^2}) \bigg] \non \\ &
- \frac{1}{m^2} \sum_{n=0} \bigg(\frac{1}{n+2}+\frac{z}{n+1}\bigg)
\left( 1-\frac{y_i}{m^2}\right)^n~.
\end{align}

For the conversion from the circular to the Cartesian basis, and 
to ensure finite results in the limit $\sin \theta \to 0$,
it is important that $a_{L+}$ and $a_{+-}$ vanish at
$\sim \sin \theta$ and $\sin^2 \theta$ respectively. This 
corresponds to the requirement that the factors in curly brackets 
do not diverge for small $\sin \theta$, i.e. in the 
limit $(m^2(1-q^2)-y_i) \to 0$. This is guaranteed by the 
behaviour of $N(y_i)$ for $y_i \rightarrow m^2 (1-q^2)$:
\begin{equation}
\frac{m^2 N(y_i)}{m^2(1-q^2)-y_i} \bigg|_{y_i \to m^2(1-q^2)}  = 
-\frac{\log(1-q^2)}{q^2}-\frac{\log(q^2)}{1-q^2}~,
\end{equation}
The results for $a^{(1,m)}_{ij}$ in the Cartesian basis are listed in 
Appendix~\ref{app:cartesian}. 

We note that the imaginary part of $L_{\mu \nu}$, which is present
in the coefficients $a_{L+}$ or $a_{-1}$ only, is of interest for 
those cases where the hadronic current receives contributions from 
different amplitudes with non-trivial relative phases. This is 
possible, e.g. for final states with three or more mesons or for 
$p\bar{p}$ production.

\section{Tests of the result}

After summation over the polarizations of the virtual photon the 
differential rate is given by the properly contracted leptonic 
tensor:
\begin{align}
& \frac{q^2}{(4 \pi \alpha/s)^2} L^{\mu \nu} (q_{\mu} q_{\nu}-g_{\mu \nu} q^2) = 
a_{LL} + 2a_{++} \non \\ 
& = -3 \; a_{00} 
+ ( \frac{1}{4q^2}(1-y_1)^2 - m^2) \; a_{11} \non \\ &
+ ( \frac{1}{4q^2}(1-y_2)^2 - m^2) \; a_{22} \non \\ &
+ ( \frac{1}{2q^2}(1-y_1)(1-y_2) - (1-2m^2)) \; a_{12} \non \\
& = 
L_0 \bigg\{1 +
\frac{\alpha}{\pi} \bigg[ -\log(4w^2) [1+\log(m^2)] \non \\ 
& - \frac{3}{2} \log(\frac{m^2}{q^2}) - 2 + \frac{\pi^2}{3} \bigg] \bigg\}
+ \frac{\alpha}{\pi} \bigg\{ 4 \bigg( 
\frac{1+(1-y_2)^2}{2y_1 y_2} L(y_1) \non \\ & 
+ \frac{1-2q^2}{2(1-q^2)^2} \log(q^2) 
+ \bigg[ 1-\frac{y_1-2y_2}{2(1-y_2)} - \frac{y_1 y_2}{4(1-y_2)^2}
\bigg] \non \\ & \times \log(\frac{y_1}{q^2}) 
- \frac{1}{2(1-q^2)} + \frac{1}{4(1-y_1)} + 
\frac{1-y_2}{4y_1} \non \\ &  + [y_1 \leftrightarrow y_2] \bigg) 
- 2 a_{++}^{(1,m)} \bigg\}~, 
\label{tensortrace}
\end{align}
where
\begin{align}
L_0 &= \frac{q^2}{(4 \pi \alpha/s)^2} 
L_0^{\mu\nu} (q_{\mu}q_{\nu}-q^2 g_{\mu\nu}) 
= \non \\ & -\frac{2}{y_1 y_2} \bigg[ 2q^2+y_1^2+y_2^2 
-\frac{2m^2q^2(1-q^2)^2}{y_1 y_2} \bigg]~,
\end{align}
$L_0^{\mu\nu}$ being the leptonic tensor at LO. In $L_0$ only 
the relevant terms in the limit $m^2 \to 0$ have been kept.
Large angle terms and mass corrections are in agreement with 
Berends {\it et al.}~\cite{Berends:1986yy,Berends:1988ab}.
After proper analytic continuation and using the substitutions
\begin{align}
t \rightarrow -y_1+i \eta~, \quad s \rightarrow -y_2+i \eta~, \quad 
u \rightarrow s+i\eta~,
\end{align}
the leptonic tensor in~\eq{generaltensor} is closely 
related\footnote{We thank N.P. Merenkov for drawing our attention to 
this reference.} to the tensor $T_{\mu \nu}$, which describes ``virtual 
Compton scattering'' and has been calculated by Kuraev 
{\it et al.}~\cite{Kuraev:cg1}. After performing this analytical 
continuation, the results are in agreement 
(However, the result for $T_{12}$ printed in~\cite{Kuraev:cg2}
contains a typo and would give rise to a discrepancy.).

\section{Conclusions}

Compact analytical formulae have been obtained for the virtual 
corrections to the process $e^+ e^- \to \gamma \gamma^*$, which 
are valid for photon emission under both large and small angles.
After proper analytic continuation the results are in agreement
with those obtained in~\cite{Kuraev:cg1} for the reaction 
$\gamma^*+e^-\to\gamma+e^-$. In polarization averaged form they are 
in agreement with those for $e^+ e^- \to \gamma Z$ 
from~\cite{Berends:1986yy,Berends:1988ab}.

\section*{Acknowledgements}

We would like to thank G.~Cataldi, A.~Denig, S.~Di~Falco, W.~Kluge, 
S.~M\"uller, G.~Venanzoni and B.~Valeriani for reminding us constantly 
of the importance of this work for the experimental analysis
and H.~Czy\.z and N.P.~Merenkov for very useful discussions.
Work supported in part by BMBF under grant number 05HT9VKB0 and 
E.U. EURO\-DA$\Phi$NE network TMR project FMRX-CT98-0169.

\appendix

\section{The leptonic tensor in the Cartesian basis}

\label{app:cartesian}

For the convenience of the reader we list the mass-suppressed 
terms $a_{ij}^{(1,m)}$ also in the Cartesian basis. The large-angle
contributions have been given in~\cite{Rodrigo:2001jr}.
The component proportional to $g_{\mu \nu}$ reads
\begin{align}
a_{00}^{(1,m)} =  - a_{++}^{(1,m)}~,
\end{align}
see \eq{apiupiu}. The coefficient $a_{+-}$ does not contribute at this 
order to $a_{00}$. The remaining components are given by
\begin{align}
a_{11}^{(1,m)} &= \frac{q^2}{1-q^2} \bigg\{
\frac{4m^2}{y_1^2} \bigg[1-\log(\frac{y_1}{m^2}) \non \\ &
+ \frac{m^2}{y_1} \bigg( Li_2(1-\frac{y_1}{m^2})-\frac{\pi^2}{6} \bigg) \bigg]
- n(y_1,1) \non \\ &
+\frac{2m^2 q^2}{y_1(m^2(1-q^2)-y_1)} \bigg[
\frac{1}{q^2} \log(\frac{y_1}{m^2}) + \frac{\log(q^2)}{1-q^2} \non \\ &
+ \bigg(1+\frac{m^2}{m^2(1-q^2)-y_1}\bigg) N(y_1) \bigg] \bigg\} + \non \\ &
+ \frac{1}{1-q^2} \bigg\{
 \frac{4m^2(1-q^2)}{y_2^2} \bigg[ \log(q^2) \log(\frac{y_2^4}{m^4 q^2}) 
\non \\ & + 4 Li_2(1-q^2) 
+ 2 \bigg( Li_2(1-\frac{y_2}{m^2})-\frac{\pi^2}{6} \bigg) \bigg] \non \\ & 
+ \frac{4m^2 q^2}{y_2^2} \bigg[1-\log(\frac{y_2}{m^2}) 
+ \bigg(1+\frac{m^2}{y_2}\bigg)
 \bigg( Li_2(1-\frac{y_2}{m^2}) \non \\ & 
 -\frac{\pi^2}{6} \bigg) \bigg] 
- \frac{1-2q^4}{q^2} n(y_2,\frac{3-8q^2+6q^4}{1-2q^4}) \non \\ &
+\frac{2m^2}{y_2(m^2(1-q^2)-y_2)} \bigg[
\frac{1}{q^2} \log(\frac{y_2}{m^2}) + \frac{\log(q^2)}{1-q^2}  \non \\ &
+ \bigg(3+\frac{m^2}{m^2(1-q^2)-y_2}\bigg) N(y_2) \bigg] \bigg\}~,
\end{align}
\begin{align}
a_{22}^{(1,m)} &= a_{11}^{(1,m)}(y_1\leftrightarrow y_2)~,
\end{align}
\begin{align}
a_{12}^{(1,m)} &= \frac{q^2}{1-q^2} \bigg\{
 \frac{4m^2}{y_1^2} \bigg[1-\log(\frac{y_1}{m^2}) \non \\ &
 + \bigg( \frac{1}{2} + \frac{m^2}{y_1} \bigg)
 \bigg( Li_2(1-\frac{y_1}{m^2})-\frac{\pi^2}{6} \bigg) \bigg] \non \\ &
- \frac{1-q^2}{q^2} n(y_1,\frac{1}{1-q^2})
+ \frac{2m^2}{y_1(m^2(1-q^2)-y_1)} \non \\ & \times \bigg[
\frac{1}{q^2}\log(\frac{y_1}{m^2}) + \frac{\log(q^2)}{1-q^2} \non \\ &
+ \bigg(2+\frac{m^2}{m^2(1-q^2)-y_1}\bigg) N(y_1)  \bigg] \bigg\} 
 + [y_1 \leftrightarrow y_2]~.
\end{align}
and
\begin{align}
a_{-1}^{(1,m)} &= 0~.
\end{align}

\section{Scalar one-loop integrals}

\label{loopintegrals}

A few two-, three-, and four-point scalar one-loop integrals
enter our calculation. Expression for the two-point scalar 
integrals are simple and well known.
The notation from~\cite{Rodrigo:2001jr}, where the corresponding results 
valid for large photon angles ($m^2 \ll 1,q^2,y_i$) can be found,
is used in the following.
The general three-point scalar one-loop integral is defined by 
\begin{align}
& C_0(p_a^2,(p_a-p_b)^2,p_b^2,m_1^2,m_2^2,m_3^2) 
= -i 16\pi^2 \mu^{4-D} \non \\ 
& \times \int \frac{d^D k}{(2\pi)^D} 
\frac{1}{[k^2-m_1^2][(k-p_a)^2-m_2^2][(k-p_b)^2-m_3^2]}~.  
\end{align}
Four different three-point scalar one-loop integrals are needed 
\begin{align}
& C01 = C_0((p_i-k_1)^2,0,m_e^2,0,m_e^2,m_e^2)~, \non \\ 
& C02 = C_0(m_e^2,s,m_e^2,0,m_e^2,m_e^2)~, \non \\
& C03 = C_0((p_i-k_1)^2,Q^2,m_e^2,0,m_e^2,m_e^2)~, \non \\
& C04 = C_0(Q^2,s,0,m_e^2,m_e^2,m_e^2)~,
\end{align}
$i=1,2$, and one scalar box 
\begin{align}
& D0  = -i 16\pi^2 \mu^{4-D} \int \frac{d^D k}{(2\pi)^D}  \\
& \times
\frac{1}{k^2[(k+p_i)^2-m_e^2][(k+p_i-k_1)^2-m_e^2][(k-p_j)^2-m_e^2]}~, \non    
\end{align}
with $j \ne i$. 
 
The following simple expressions are used, from where the limits
($m^2 \ll 1,q^2,y_i$) or ($m^2 \ll 1, q^2$ but $m^2 \sim y_i$)
are obtained:
\begin{align}
& C01 = 
\frac{s^{-1}}{y_i} \bigg[Li_2(1-\frac{y_i}{m^2}) - \frac{\pi^2}{6}
\bigg]~, \non \\
& C02 = \frac{s^{-1}}{\beta} \bigg[ 
\bigg(\Delta -2\log(\beta) -\frac{\log(c)}{2} \bigg) \log(c) \non \\ & \quad
- 2 Li_2(c) - \frac{2\pi^2}{3}
+ i \pi \bigg(\Delta -2 \log(\beta) \bigg)\bigg]~, \non \\
& C04 = \frac{s^{-1}}{1-q^2} \bigg[\frac{\log(c)}{2} - \frac{\log(c_q)}{2}
+ i \pi \log(\frac{c}{c_q})\bigg]~, \non \\
& D0 =  \frac{s^{-2}}{\beta y_i}
\bigg[ - \bigg( \Delta + 2 \log (\frac{m}{y_i}) \bigg) \log(c_q) +\log^2(c) 
\non \\ & \quad 
+ 2 Li_2(1-\frac{c_q}{c}) +2 Li_2(1-c\; c_q) -Li_2(1-c_q^2) - \pi^2 
\non \\ & \quad - i \pi \bigg( \Delta + 2 \log (\frac{m}{c \; y_i}) \bigg)
\bigg]~,
\end{align}
with
\begin{equation}
\Delta = \frac{(4\pi)^\e}{\e \; \Gamma(1-\e)} 
\left( \frac{\mu^2}{s} \right)^{\e}~,
\label{delta}
\end{equation}
and
\begin{align} 
\beta &= \sqrt{1-4m^2}~, \qquad c =\frac{1-\beta}{1+\beta}~, \non \\
\beta_q &= \sqrt{1-\frac{4m^2}{q^2}}~, \qquad c_q =\frac{1-\beta_q}{1+\beta_q}~.
\end{align}
Our expression for the C03 function is rather cumbersome:
\begin{align}
C03 &=\frac{s^{-1}}{q^2(z_1-z_2)}\bigg[\log(\frac{y_i}{q^2}) 
\log\left(\frac{(1-z_1)z_2}{(1-z_2)z_1}\right) \non \\ &
+ \bigg\{Li_2(\frac{1}{z_1})
- \log((z_3-z_1)(z_4-z_1)) \log(1-\frac{1}{z_1}) \non \\ &
+ Li_2(\frac{1-z_1}{z_3-z_1}) - Li_2(\frac{-z_1}{z_3-z_1}) \non \\ &
+ Li_2(\frac{1-z_1}{z_4-z_1}) - Li_2(\frac{-z_1}{z_4-z_1}) 
- [z_1 \leftrightarrow z_2] \bigg\} \bigg]~,
\end{align}
where
\begin{align}
z_{1,2}&=\frac{q^2+y_i}{2q^2}\left( 1 \pm 
\sqrt{1-\frac{4(m^2-i \eta)q^2}{(q^2+y_i)^2}}\right)~, \non \\ 
z_{3,4}&=\frac{1}{2}\left( 1 \pm 
\sqrt{1-\frac{4(m^2-i \eta)}{q^2}}\right)~,
\end{align}
being 
\begin{align}
Im(C03) = \frac{\pi}{q^2(z_1-z_2)} \log 
\left( \frac{(z_3-z_1)(z_4-z_2)}{(z_4-z_1)(z_3-z_2)} \right)~,
\end{align}
its imaginary part. As for the other scalar one-loop functions, 
this expression is expanded up to terms of order $m^2$ and $m^4$.


\end{document}